\documentclass[aps,prl,twocolumn,superscriptaddress,showpacs,amsmath]{revtex4-1}
\usepackage{graphicx,hyperref}
\begin{document}

\title{Thermally assisted ordering in Mott insulators}
\author{Hunter Sims}
\affiliation{Computational Materials Science, German Research School for Simulation Sciences, 52425 J\"ulich, Germany}
\author{Eva Pavarini}
\affiliation{Institute for Advanced Simulation, Forschungszentrum J\"ulich, 52425 J\"ulich, Germany}
\affiliation{JARA High-Performance Computing, 52425 J\"ulich, Germany}
\author{Erik Koch} \email{e.koch@fz-juelich.de}
\affiliation{Computational Materials Science, German Research School for Simulation Sciences, 52425 J\"ulich, Germany}
\affiliation{Institute for Advanced Simulation, Forschungszentrum J\"ulich, 52425 J\"ulich, Germany}
\affiliation{JARA High-Performance Computing, 52425 J\"ulich, Germany}

\date{\today}
\begin{abstract}
Ginzburg-Landau theory describes phase transitions as the competition between energy and entropy: 
The ordered phase has lower energy, while the disordered phase has larger entropy. 
When heating the system, ordering is reduced entropically until it vanishes at the critical temperature. This established picture implicitly assumes that the energy difference between ordered and disordered phase does not change with temperature. 
We show that for the Mott insulator KCuF$_3$ this assumption is strongly violated: thermal expansion energetically stabilizes the orbitally-ordered phase to such and extent that no phase transition is observed. 
This new mechanism explains not only the absence of a phase transition in KCuF$_3$ but even suggests the possibility of an {\em inverted} transition in closed-shell systems, where the ordered phase emerges only at high temperatures. 
\end{abstract}

\maketitle

Mott insulators with orbital degrees of freedom often exhibit  orbitally ordered phases \cite{Tokura2000}.
There are two established explanations for this:
(i) electron-phonon coupling induces cooperative Jahn-Teller distortions \cite{Kanamori1960} that lead to orbital ordering or 
(ii) Kugel-Khomskii superexchange \cite{KK1973} gives rise to orbital order that leads to a cooperative 
lattice distortion. 
Since both mechanisms tend to result in the same type of ordering, identifying which one drives it is a `chicken-and-egg problem' \cite{KhomskiiBook}. 
Even though they strongly differ, these two mechanisms have one fundamental aspect in common: Their hallmark
is a conventional Ginzburg-Landau-type transition \cite{ToledanoBook} between a low-temperature ordered structure and a symmetric high-temperature phase \cite{BersukerBook}.
Here we show that this conventional picture overlooks a key aspect: Taking thermal expansion into account
leads to a novel scenario, and to a third, unconventional, ordering mechanism.
We find that, as consequence, the order is not necessarily destroyed by temperature. 
In fact, for the prototypical orbital-ordering perovskite KCuF$_3$ \cite{KK1973} we explain how thermal expansion 
favors the symmetry-broken phase with an order parameter that {\em increases} with temperature. 
The key feature of the new mechanism is the strong dependence 
of the energy gained by breaking the symmetry
on the lattice constant, and ultimately, via thermal expansion, on temperature.
We anticipate that this thermally-assisted ordering can operate even in closed-shell systems. 
This would result in an inverted Ginzburg-Landau transition, with symmetry-breaking {\em above} a critical temperature. 
These surprising conclusions are based on very general arguments. 
We thus expect that similar effects will play a key role in other ordering phenomena of totally different nature.

\begin{figure}
\center
\includegraphics[width=0.65\columnwidth]{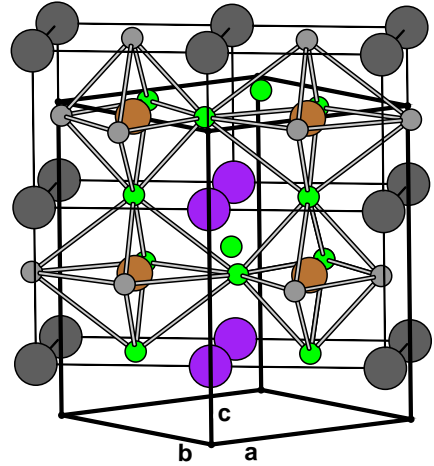} 
\caption{\label{fig:crystal}(Color online) Crystal structure of KCuF$_3$. Inequivalent atoms inside the I4/mcm unit cell (thick black lines) are shown in color (Cu: brown, F: green, K: violet). The additional atoms in grey show the pseudocubic setting in which the network of corner sharing octahedra becomes apparent. 
The pseudocubic axes are defined as $\mathbf{x}=(\mathbf{a}+\mathbf{b})/2$, $\mathbf{y}=(-\mathbf{a}+\mathbf{b})/2$, and $\mathbf{z}=\mathbf{c}/2$. For clarity lattice distortions are exaggerated twofold.}
\end{figure}

Following the seminal work of Kugel and Khomskii \cite{KK1973}, the fluoride KCuF$_3$ 
is considered  the prototype of an orbitally ordered material. 
Its structure \cite{Okazaki61}, shown in Fig.~\ref{fig:crystal}, derives 
from a cubic perovskite with Cu in $d^9$ configuration with one hole in the $e_g$ orbital 
surrounded by an octahedron of fluoride ions in a cage of potassium ions. The actual crystal 
shows a tetragonal compression, slightly lifting the $e_g$ degeneracy.
The fluoride ions in the $a$-$b$ plane move along the lines connecting the Cu ions 
such that long ($\ell$) and short ($s$) bonds alternate in the $x$ and $y$ directions, 
leading to a cooperative Jahn-Teller distortion and a competing splitting of the 
$e_g$ orbital. The distortion pattern also alternates in the $z$ direction,
resulting in an antiferrodistortive orbital-ordering.
The tetragonal distortion is parametrized by 
$c/a\sqrt{2}$ and the Jahn-Teller distortion by $\delta=(\ell-s)/a\sqrt{2}$.

\begin{figure}
 \center
 \includegraphics[width=\columnwidth]{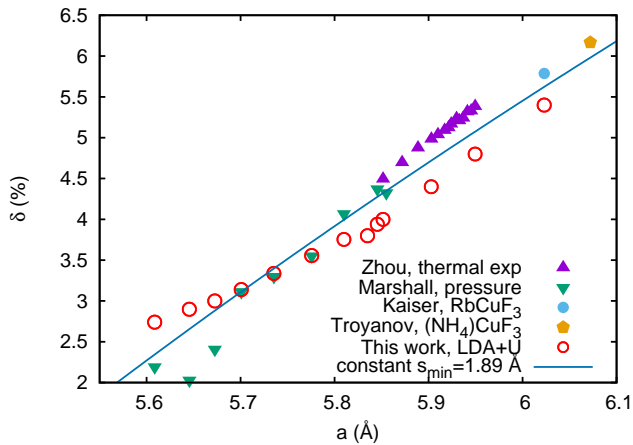} 
 \caption{\label{fig:deltavsa}(Color online) Distortion parameter $\delta$ as a function of lattice constant $a$ in 
 thermally expanding KCuF$_3$ \cite{Marshall2013} and under hydrostatic pressure \cite{Zhou2011}, 
 for RbCuF$_3$ \cite{Kaiser1990}, and (NH$_4$)CuF$_3$ \cite{Troyanov1993} compared to our 
 calculations and the values obtained for constant short Cu--F distance $s_\mathrm{min}$.}
\end{figure}

The mechanism driving the distortion $\delta$ and orbital-ordering has been the subject of 
intense controversy. As early as 1960, Kanamori noted that the structure of KCuF$_3$ 
could arise from the cooperative Jahn-Teller effect \cite{Kanamori1960}. Later, 
Kugel and Khomskii showed that orbital order in KCuF$_3$ can originate from electronic superexchange even in the absence of distortions \cite{KK1973}. As one of the first applications 
of the density-functional theory plus $U$ method (DFT+U) \cite{LDAU}, Liechtenstein \textit{et al.}\ found that a Hubbard $U$ 
is necessary to stabilize the distorted structure and concluded that an electronic 
Kugel-Khomskii mechanism drives the transition. In the same year 
Towler \textit{et al.}\ found that Hartree-Fock also gives reasonable agreement with 
experiment, despite the complete lack of correlations \cite{Towler1995}. Eventually, 
the question was settled by density-functional theory plus dynamical mean-field theory (DFT+DMFT) calculations, which showed that 
Kugel-Khomskii superexchange alone can only account for orbital order below  
$T_\mathrm{KK}\approx 350$~K \cite{KCuF3}, while experimentally it persists to much
higher temperatures \cite{Okazaki61}. 
In fact, the transition to the undistorted high-temperature phase has never been 
seen experimentally, and the analysis of the crystal structure up to 900~K showed that, 
contrary to conventional wisdom, the distortion {\em increases} with temperature 
instead of vanishing above some critical value \cite{Marshall2013}. Applying hydrostatic 
pressure dramatically reduces the distortion as the lattice constant decreases \cite{Zhou2011},
while expanding the lattice by chemical pressure, substituting K by Rb \cite{Kaiser1990} or 
NH$_4$ \cite{Troyanov1993}, results in an increase of the distortion parameter $\delta$ 
following the same trend as in thermally expanded KCuF$_3$.
All this points to the lattice constant as the key player in determining the 
size of the distortion (see Fig.~\ref{fig:deltavsa}). 
Remarkably, the importance of this  has been missed so far.

To understand the role of the lattice, we have performed DFT+U calculations for KCuF$_3$ with the experimental lattice constants at different temperatures. 
Calculations were performed using the Vienna \textit{ab-initio} Simulation Package ({\sc vasp}) \cite{vasp1} within the generalized gradient approximation (GGA) of Perdew, Burke, and Ernzerhof \cite{PBE} to density functional theory 
using the projector augmented-wave (PAW) \cite{PAW} pseudopotentials of Kresse and Joubert \cite{pseudo}. We apply onsite Coulomb interactions on the Cu 3$d$ orbitals through the ``+$U$'' correction of Liechtenstein \textit{et al.} \cite{LDAU} with double-counting corrections in the fully localized limit. 

\begin{figure}
\center
\includegraphics[width=\columnwidth]{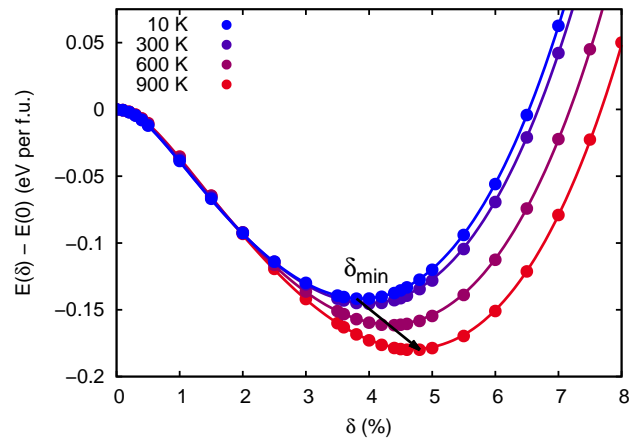} 
\caption{\label{fig:delta}(Color online) DFT+U energy gain per formula unit as a function of the distortion parameter $\delta$ for experimental unit cells at different temperatures. With thermal expansion the minimum of the energy curve moves to larger distortions $\delta_\mathrm{min}$ and deepens. Lines are fits to guide the eye.}
\end{figure}

It is known that DFT+U describes structural properties remarkably well \cite{Binggeli2004}.
We find that also the energy gained by moving the fluorine ions agrees with both experimental estimates \cite{Ghigna2010} 
and calculations explicitly including many-body effects \cite{Leonov2008,Flesch2012}.
Moreover, extracting the frequency of the $A_{1g}$ mode, we find excellent 
agreement with Raman data \cite{Abbamonte2012}. 
Our results are fairly independent of the model parameters $U$ and $J$, as long as they 
are large enough to open a gap.  
The main effect of increasing $U$ is to slightly increase the effective radius of the 
cation \cite{Autieri2014}.
Fig.~\ref{fig:delta} shows the energy gained by distorting the lattice for the experimental 
unit cell parameters at increasing temperatures \cite{Satija1980,Marshall2013} calculated 
using the established values $U=7$~eV and $J=0.9$~eV \cite{LDAU}. 
We find that the distortion $\delta$
increases with lattice constant in good agreement with the experimental values 
(see Fig.~\ref{fig:deltavsa}). 

While our calculations reproduce the observed increase of the distortion very well,
neither of the established theoretical mechanisms can explain it:
The orbital superexchange coupling 
decreases with distance so that the Kugel-Khomskii mechanism weakens as the lattice 
expands \cite{KK1973}. The Jahn-Teller mechanism could in principle explain a 
distortion that increases with volume. Writing the energy gained by displacing the 
fluorine ions by $\Delta=(\ell-s)/2=a\,\delta/\sqrt{2}$ from their symmetric position 
as $E_\mathrm{JT}(\Delta)=-g\Delta+C\Delta^2/2$, where $g$ gives the splitting of the 
$e_g$ level and $C$ the elastic constant, the energy is minimized for 
$\Delta_\mathrm{JT}=g/C$ \cite{Kanamori1960}. 
The dependence of $\Delta_\mathrm{JT}$ 
on the lattice constant $a$ is thus given by the change of $g$ and $C$. 
Both will decrease with $a$, and if $C$ decreases much faster than $g$, 
$\Delta_\mathrm{JT}$ could become arbitrarily large. From crystal-field theory 
we know that $g$ scales with $1/a^4$. To obtain the experimentally observed increase 
in $\Delta_\mathrm{min}$ the elastic constant $C$ would have to decay faster 
than $1/a^{14}$. This 
contradicts, however, the observed temperature (volume) dependence of the 
$A_{1g}$-mode \cite{Abbamonte2012}, ruling out the Jahn-Teller mechanism as well.

\begin{figure}
\center
\includegraphics[width=\columnwidth]{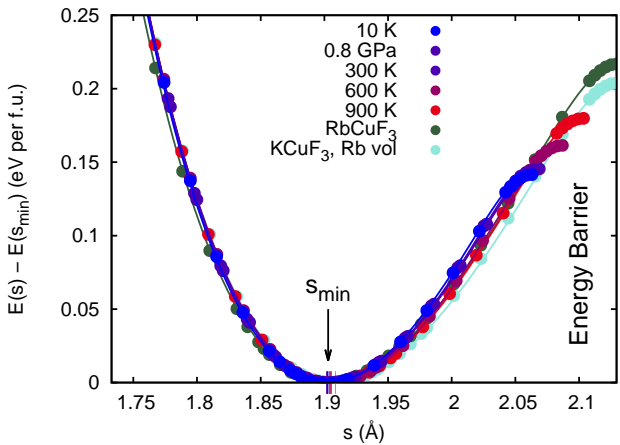} 
\caption{\label{fig:short}(Color online) Change in DFT+U energy as a function of Cu--F distance $s$ for different lattice constants $a$. For $s$ smaller than $s_\mathrm{min}$ the energy curves are practically independent of the actual lattice. For $s$ larger than $s_\mathrm{min}$ each curve reaches a maximum at the undistorted position $s=a/2\sqrt{2}$.}
\end{figure}

The complete failure of the established distortion mechanisms calls for a 
change of perspective. Instead of focussing on the displacement from the 
high-symmetry position, we consider the Cu--F distance. The shortest distance $s$ should be
given by when the ions touch.
In fact, for KCuF$_3$ it is practically independent of temperature \cite{Marshall2013}
so that the increase of $\Delta$ is simply a consequence of thermal expansion. 
The same is true when applying pressure \cite{Zhou2011}, substituting K by Rb \cite{Kaiser1990}, or NH$_4$ \cite{Troyanov1993}.
We can make this picture quantitative by plotting the energy curves of Fig.~\ref{fig:delta} 
as a function of the Cu--F distance (see Fig.~\ref{fig:short}): pushing the ions closer 
together than the optimal distance results in a strong repulsion --- 
a Born-Mayer potential \cite{BornMayer}. 

This suggests a straightforward model: we describe $E(\Delta)$ using a Born-Mayer 
repulsion energy $E_\mathrm{BM}=A\exp(-r_{\mathrm{Cu-F}}/\rho)$, the Ewald energy $E_\mathrm{Ewald}$ of the periodic arrangement 
of ions, and a term $E_\mathrm{CF}=\Gamma-\sqrt{\Gamma^2+(g\Delta)^2}$ describing the crystal-field splitting of the 
$e_g$ level due to the tetragonal compression and the displacement 
$\Delta$. Since the ionic charges are practically independent of the lattice constant, 
as are the Born-Mayer coefficients $A$ and $\rho$, the couplings $\Gamma$ and $g$ are the 
only parameters that depend on the lattice constant.
As expected, the crystal-field splitting scales as $1/a^4$, 
while the change in Ewald energy can be approximated by $-C_\mathrm{Ewald}\,\Delta^2/2$, where 
$C_\mathrm{Ewald}$ scales as $1/a^3$. 
The resulting expression 
\begin{equation*}
 \begin{split}
 E_\mathrm{ionJT}(\Delta) =\; & \Gamma-\sqrt{\Gamma^2-(g\Delta)^2} 
                          - C_\mathrm{Ewald}\Delta^2/2\\
                          &\;\;+ 2Be^{-a/\rho 2\sqrt{2}}(\mathrm{cosh}(\Delta/\rho)-1)
 \end{split}
\end{equation*}
gives not only excellent fits to the DFT+U energies for KCuF$_3$ as shown by the curves in 
Figs.~\ref{fig:delta} and \ref{fig:short} with the parameters given in Table~\ref{stab:fitpar}, 
but should also describe strongly ionic Jahn-Teller-active compounds in general.
\begin{table}
 \center
 \begin{tabular}{r|rr|rrr}
  $T$ (K) & $a$ (\AA) & $c$ (\AA) & $\Gamma$ (eV) & $g$ (eV/\AA) & $C_\mathrm{Ewald}$ (eV/\AA$^2$)\\
  \hline
   10 & 5.835 & 7.828 & 0.0620 & 2.195 & 20.059\\ 
  300 & 5.852 & 7.841 & 0.0640 & 2.173 & 19.877\\ 
  600 & 5.903 & 7.897 & 0.0677 & 2.112 & 19.342\\
  900 & 5.950 & 7.954 & 0.0698 & 2.060 & 18.883
 \end{tabular}
 \caption{\label{stab:fitpar}Model parameters for $E_\mathrm{ionJT}(\Delta)$
 that fit the DFT+U curves calculated for KCuF$_3$ with experimental 
 lattice constants at different temperatures (Figs.~\ref{fig:delta} and \ref{fig:short}).
 $\Gamma$ increases with the tetragonal distortion, while $g$ decreases with $a$. 
 The Born-Mayer parameters $B=9188$~eV and $\rho=0.2186$~\AA\ are independent of the lattice constants.
 This is also true for the charges of the ions entering the Ewald energy: 
 $Z_\mathrm{Cu}=1.86\,e=2Z_\mathrm{K}=-2Z_\mathrm{F}$. 
}
\end{table}

We can now explain the anomalous behavior seen in KCuF$_3$.
To simplify the discussion we neglect for the moment the tetragonal splitting, setting $\Gamma=0$.
The Cu$^{2+}$ cation is fairly small, i.e., $\rho\ll a/2\sqrt{2}$, so that the frequency about the undistorted position, $m\omega_0^2=\left.d^2 E(\Delta)/d\Delta^2\right|_{\Delta=0}=2Be^{-a/\rho 2\sqrt{2}}/\rho^2-C_\mathrm{Ewald}$, is quite low. In a simple Jahn-Teller picture this would imply a very large distortion $\Delta_\mathrm{JT}$, that would bring Cu and F extremely close to each other. 
In reality, however, the ions repel strongly at short distance. Since this Born-Mayer repulsion increases exponentially, the distortion will be stopped at a Cu--F distance $s_\mathrm{min}$ that is practically independent of the lattice constant. 
The observed linear increase of the distortion with the lattice constant $\Delta_\mathrm{min}(a)\approx a/2\sqrt{2}-s_\mathrm{min}$ is thus simply the consequence of a constant $s_\mathrm{min}$
(see the line in Fig.~\ref{fig:deltavsa}).
 At the same time the energy gained from the distortion increases with $\Delta_\mathrm{min}$. The large thermal expansion $a(T)$ thus stabilizes the distortion in KCuF$_3$, explaining the absence of a 
transition to the undistorted structure. 
We note that in our model the frequency $\omega_0$ differs from the frequency of the $A_{1g}$ Raman mode, which is given by the expansion about the minimum: $m\omega^2_{A_{1g}}=\left.d^2E(\Delta)/d\Delta^2\right|_{\Delta=\Delta_\mathrm{min}}$. The difference is due to the  Born-Mayer potential, which makes the $A_{1g}$-mode quite anharmonic, in agreement with experiment \cite{Abbamonte2012}.
The scenario of fixed $s_\mathrm{min}$ is not limited to KCuF$_3$. In fact, Table~5 of Ref.~\cite{Kaiser1990} lists the short Cu--F distances $s$ of thirteen materials of widely varying structure and composition with Jahn-Teller-active CuF$_6$ octahedra. They all differ by less than 2\%.

\begin{figure}
 \center
 \includegraphics[width=\columnwidth]{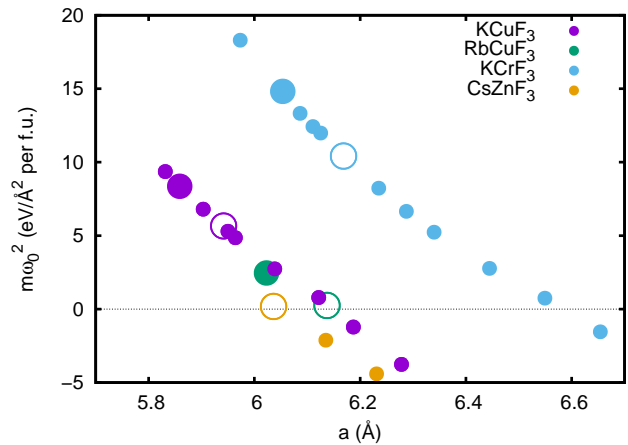} 
 \caption{\label{fig:landau}(Color online) Dependence of the calculated elastic constant $m\omega_{0}^{2}$ for the distortion about $\Delta=0$ on the lattice parameter $a$. For the open-shell systems the values are calculated by DFT+U with $U=7$~eV and $J=0.9$~eV for KCuF$_3$ and RbCuF$_3$, and $U=6$~eV and $J=0.9$~eV for KCrF$_3$, while the closed shell CsZnF$_3$ is already gapped without a Hubbard $U$. Large filled circles indicate the experimental lattice constant at room temperature, open circles the DFT+U or DFT relaxed lattice constant, which are between 1--2~\% larger than the experimental values. 
 For all compounds the elastic constant changes sign when the lattice constant gets large enough. The larger the cation, the larger the critical lattice constant $a_c$. For the smallest, Zn, the relaxed structure is tantalizingly close to the value required for an inverted Ginzburg-Landau transition.
}
\end{figure}

For larger cations, $\omega_0$ will be harder, leading to smaller distortions and a more Jahn-Teller-like picture. There is, however, a crucial difference: $\omega_0$ 
softens dramatically with the expansion of the lattice, leading to a robust distortion 
even as the temperature increases. Remarkably, this is what is actually observed in the 
tetragonal phase of another
fluoride, KCrF$_3$, up to the volume-collapse transition at 973 K \cite{Margadonna2007}. 
The persistence of the distortion with increasing temperature 
is expected for all strongly ionic Jahn-Teller-active compounds with significant 
thermal expansion coefficients. This thermally assisted ordering mechanism should be
particularly useful for engineering materials, where we want the symmetry-broken phase to survive to 
high temperatures \cite{Goodenough2013}.
Moreover, it suggests an intriguing scenario: When $a$ exceeds the critical value 
$a_c=\rho\,2\sqrt{2}\ln(2B/\rho^2 C_\mathrm{Ewald}(a_c))$ the frequency $\omega_0$ becomes imaginary so that even a system with a non Jahn-Teller-active cation would start to distort. When $a_c$ is crossed in thermal expansion such a system 
could show an {\em inverted} Ginzburg-Landau transition from a high-symmetry phase at low 
temperatures to an ordered high-temperature structure. 
Ideal candidates are compounds with large lattice constant and small B-site cation as shown in Fig.~\ref{fig:landau}.
While it might be difficult to find a material where $a_c$ can be reached by thermal 
expansion alone, it is conceivable to additionally increase the 
lattice constant by strain \cite{Haeni2004} or negative pressure \cite{Wang2015} 
to just beyond the critical value, so that this unusual phase transition can be reached.

It turns out, then, that Ginzburg-Landau theory is oversimplified in that it assumes a temperature-independent electronic Hamiltonian.
We have identified a striking example which highlights the failure of this standard model of symmetry breaking: in the Mott insulator KCuF$_3$ orbital-ordering is {\em stabilized} by thermal expansion.
In fact, this is just an instance of a more general principle: When the effective Hamiltonian describing symmetry breaking has a significant temperature dependence, 
we can expect to observe more exotic phenomena than predicted by 
Ginzburg-Landau theory.

\begin{acknowledgments}
This work has been supported by the Deutsche Forschungsgemeinschaft through FOR 1346.
Calculations have been done on JURECA at the J\"ulich Supercomputer Centre under grant GRS300. 
\end{acknowledgments}

\end{document}